\documentclass[11pt]{article}
\usepackage[margin=1in]{geometry}
\usepackage{graphicx}
\usepackage{booktabs}
\usepackage{amsmath}
\usepackage[hidelinks,hypertexnames=false]{hyperref}
\usepackage{microtype}
\usepackage{float}
\usepackage{mathptmx}

\title{When Does Distributed AI Inference Need More Wide-Area Bandwidth?\\
\large A Co-Design Evaluation of Optical, Packet, and Software Levers}
\author{Prasanna C\\Lightstorm\\\texttt{c.prasanna@lightstorm.net}}
\date{August 2026}

\begin{document}
\maketitle

\begin{abstract}
Wide-area bandwidth per unit of GPU compute falls every hardware generation: in compute-intensity-ratio terms (CIR, bytes per FLOP), the gap between on-package memory and the conventional WAN is four to five orders of magnitude, widening at roughly 12--19\% per year. Position papers---including our own earlier work---argued that this makes elastic optical wide-area capacity necessary for cross-site AI inference. Reviewers correctly objected that such arguments show more bandwidth \emph{helps}, not that provisioned bandwidth beats the alternatives: KV recomputation, cache compression, locality-aware routing, scheduling, or an overprovisioned packet backbone. This paper does the comparison. We derive a workload model predicting when moving inference state across sites beats recomputing it---a context-independent crossover at 74--111\,Gbps per stream for a 70B multi-head-attention model, falling to 9--14\,Gbps under grouped-query attention at 1/8 KV heads---and quantify five sensitivity axes: context length, attention architecture, concurrency (queueing), agentic multi-step compounding, and loss/jitter-induced effective-bandwidth collapse. We bound the economics: at list GPU prices recomputation is cheaper than transfer; transfer wins when GPU scarcity and KV reuse multiply the effective GPU cost by roughly 5--20$\times$, and modern attention moves the breakeven an order of magnitude in transfer's favour. We position the network levers correctly rather than adversarially: packet networks allocate lit capacity at millisecond timescales; optical fungibility changes the amount of lit capacity at minute timescales, substituting for packet overprovisioning economically rather than functionally. Finally, we specify a ten-metric measurement plan on a three-site production-fibre testbed and frame it as an open ecosystem exercise: no single company can---or should---assemble the end-to-end evidence alone. Every claim is bounded by the regime in which it holds; several findings weaken the naive version of our own thesis, and we state them.
\end{abstract}

\section*{About this paper}
This paper consolidates an operator's interconnect analysis after a full review cycle: two academic submissions in 2026 whose seven reviewers validated the core claim while correctly demanding that it be proven against software alternatives rather than argued. Everything analytical is backed by a reproducible model with falsifiable predictions; a ten-metric measurement plan is defined on a new deep-buried production fibre corridor, with a pilot window in late 2026 deliberately open to industry collaborators. The frame is the GPU-to-GPU continuum---accelerator $\rightarrow$ HBM $\rightarrow$ NVLink/scale-up $\rightarrow$ scale-out fabric $\rightarrow$ scale-across WAN $\rightarrow$ storage---in which the industry has invested at every tier except one. This paper supplies the operator's quantified case for that missing tier: when it binds, when it does not, and what it must guarantee.

\section{Introduction: the gap is real, the precision matters}
Two facts anchor this work. Throughout, we write \textbf{CIR} (compute intensity ratio) for the bandwidth a tier can deliver to a GPU divided by that GPU's dense compute rate---bytes per FLOP. We report it in \textbf{mB/F}, millibytes per FLOP, so 1\,mB/F $= 10^{-3}$\,B/F; for scale, McCalpin's classical STREAM balance is $\sim$1{,}000\,mB/F and GEMM-adjusted LLM balance is $\sim$3\,mB/F, which is where H100 on-package memory sits. CIR is a supply-side property of a tier and must not be confused with the demand-to-link ratios used in \S1--\S2; the paper keeps the two apart throughout.

The structural fact: GPU compute grows $\sim$48\%/yr (dense BF16, A100 2020 $\rightarrow$ B200 2025; the figure is identical if sparsity-enabled numbers are used consistently for both endpoints) while network capacity grows $\sim$25--32\%/yr, so bytes-per-FLOP available to a GPU falls $\sim$11--16\%/yr at every layer beyond the package---equivalently, the compute-to-bandwidth gap widens $\sim$12--19\%/yr. In CIR terms (H100 at 989 TFLOPS dense BF16; all compute figures in this paper are dense, and sparsity-enabled numbers would double every one of them): HBM 3.39\,mB/F, NVLink 0.910, scale-out InfiniBand 0.006--0.050, conventional shared WAN $1.58\times10^{-5}$--$1.58\times10^{-4}$ (1--10\,Gbps per 8-GPU node)---four to five orders of magnitude top to bottom. Down the continuum, capital has followed every tier---HBM stacks, NVLink/NVSwitch, 400/800G scale-out fabrics---except the wide-area tier, which still runs on infrastructure sized for pre-AI traffic.

The demand fact, stated precisely: a 70B multi-head-attention model at 4K context carries a 10.7\,GB KV cache; delivering it inside a 40\,ms time-to-first-token (TTFT) budget requires 2{,}147\,Gbps of transfer capacity (propagation excluded; \S6 returns to what distance does to this budget)---215$\times$ a dedicated 10\,Gbps link ($\sim$2 orders of magnitude) and $\sim$429$\times$ the equivalent asynchronous training demand ($\sim$5\,Gbps, an assumed gradient-exchange rate rather than a measured one). These are different quantities---a demand-to-link ratio and a bytes-per-FLOP ratio---and the paper keeps them separate throughout. And 2{,}147\,Gbps is the MHA bound: grouped-query attention (GQA, 1/8 KV heads) needs 268\,Gbps; latent-attention architectures $\sim$58\,Gbps~\cite{deepseek,deepseekhw}. The question this paper answers: given software and architectural mitigations, and given that packet networks are themselves agile---when, precisely, does cross-site AI inference need more provisioned wide-area bandwidth, and what is the cheapest way to supply it?

\section{Workload model: when state must cross sites}
\subsection{Transfer vs recompute---a context-independent crossover}
Disaggregated inference either moves the KV cache to the decode site or recomputes the prefill locally. Recompute costs $T_{\mathrm{prefill}} = 2 \cdot \mathrm{params} \cdot \mathrm{context} / (\mathrm{TFLOPS} \cdot \mathrm{MFU})$; transfer costs $\mathrm{KV_{bytes}} \cdot 8 / \mathrm{bandwidth}$ plus propagation. Both scale linearly in context, so the crossover bandwidth is context-independent: transfer wins above 74\,Gbps per stream (70B MHA at 50\% MFU; 111\,Gbps at 75\%), 40--60\,Gbps for 405B, 130--194\,Gbps for 8B. Below the crossover, recomputation is the right engineering answer and no network claim survives.

\subsection{Attention architecture moves the crossover an order of magnitude}
\begin{table}[H]
\centering\small
\caption{Attention architecture, cross-site demand, and the transfer-vs-recompute crossover (70B geometry at 4K context, dense BF16). The MLA row uses DeepSeek-V3's measured 70.272\,KB/token~\cite{deepseekhw}; that is a 671B model, so its crossover is not comparable to the 70B rows.}
\label{tab:attention}
\begin{tabular}{lllll}
\toprule
attention & KV/token & KV @4K & demand @40ms TTFT & crossover @50\% MFU\\
\midrule
MHA & 2.62 MB & 10.74 GB & 2{,}147 Gbps & 74 Gbps\\
GQA 1/4 & 0.66 MB & 2.68 GB & 537 Gbps & 18.5 Gbps\\
GQA 1/8 & 0.33 MB & 1.34 GB & 268 Gbps & 9.3 Gbps\\
MLA (70.3 KB/token~\cite{deepseekhw}) & 0.07 MB & 0.29 GB & 58 Gbps & n/a---different model\\
\bottomrule
\end{tabular}
\end{table}

\begin{figure}[H]
\centering
\includegraphics[width=0.85\textwidth]{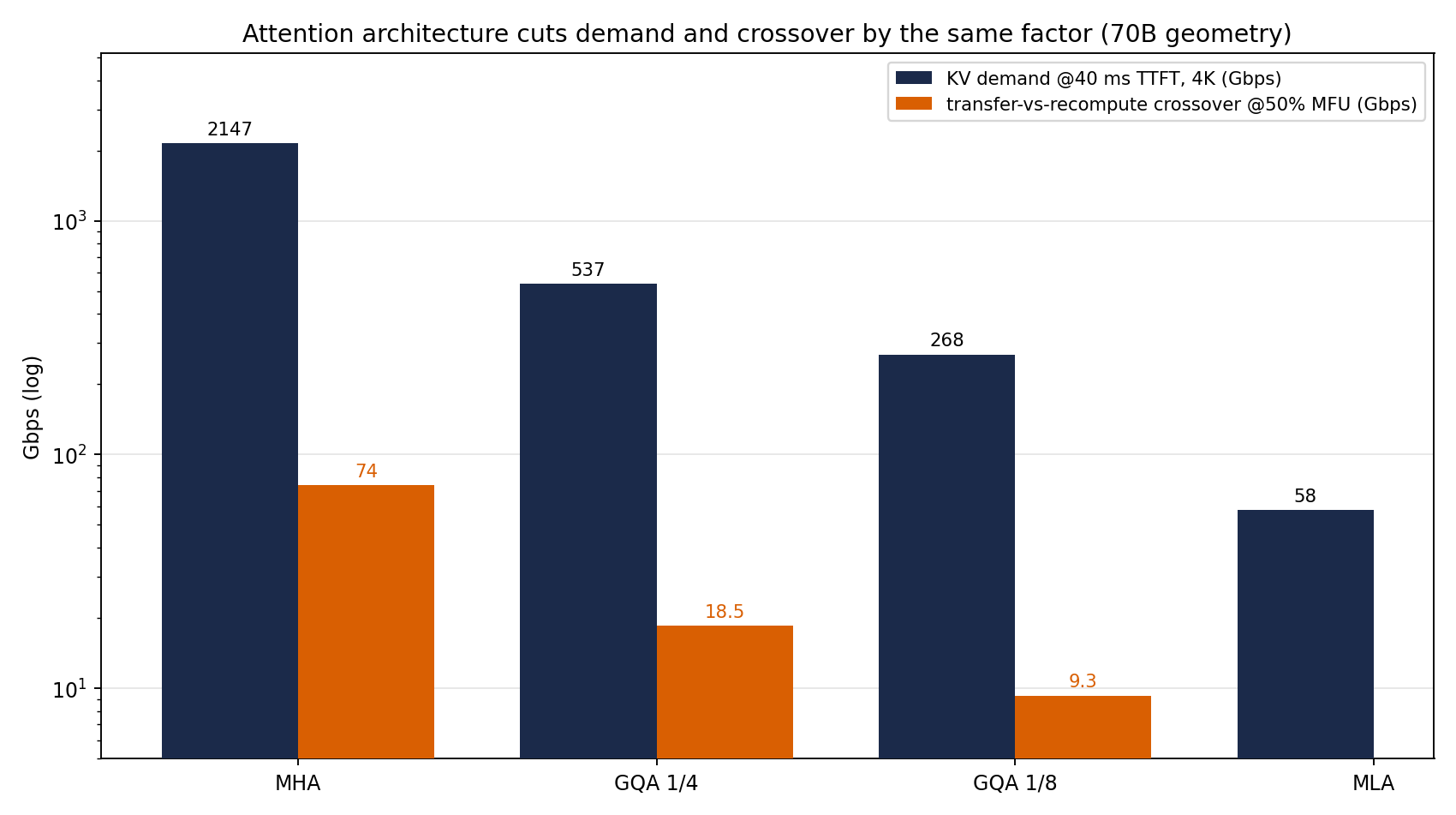}
\caption{Attention architecture cuts interactive-budget demand and the transfer-vs-recompute crossover by the same factor (70B geometry; log scale).}
\label{fig:attention}
\end{figure}

GQA and latent attention (MLA) cut cross-site demand 8--37$\times$: short-context modern-attention serving does not need heroic pipes, and a bandwidth argument that leads with the MHA number overstates its case. But the same reduction cuts the crossover to 9--14\,Gbps: link rates wide-area services provision routinely today. The net effect is not that bandwidth stops mattering; it is that cross-site disaggregation stops being exotic. State transfer becomes rational on ordinary links, moving the binding constraint from per-stream feasibility to aggregate capacity (\S2.4) and tail behaviour (\S3.5). Table~\ref{tab:attention} gives the geometry; Figure~\ref{fig:attention} shows both effects falling together.

\subsection{Locality---when state need not cross at all}
Aggregate cross-site demand $\approx$ per-request KV $\times$ request rate $\times$ (1 $-$ local hit rate). Prefix caching, session stickiness, and locality-aware routing collapse demand where request mixes are cacheable. The bandwidth case is strongest where locality is structurally low: disaggregated prefill/decode pools, multi-tenant serving, sovereignty-split workloads, and agentic pipelines hopping between specialist models at different sites.

\subsection{Agentic compounding---the strongest long-haul case}
An agent session of $S$ steps with growing context (2K + 1K/step), re-transferring accumulated KV on a fraction $f$ of site-hops, moves $f \cdot \sum \mathrm{KV}(c_i)$ bytes---roughly quadratic in $S$. Under GQA-1/8 at $f{=}0.3$, a 20-step session moves $\sim$25\,GB; ten sessions/s $\approx$ 2\,Tbps of sustained cross-site demand. Agentic serving is latency-tolerant per step but bandwidth-hungry in aggregate---bandwidth-bound even at continental distance, since relaxed per-step deadlines keep propagation negligible against transfer time.

\section{The design space: the levers, correctly positioned}
\subsection{Compression substitutes for bandwidth---up to a quality ceiling}
8$\times$ KV compression lifts a 10G link from 12\% to 52\% prefill-GPU utilisation; compression is a genuine bandwidth substitute and the design space must credit it. The limit is answer quality: ablating cached KV state drops AIME24 accuracy from 66.1\% to 50.8\% in recent measurements~\cite{memento}---the KV channel carries reasoning capacity, not just latency savings. Compression buys headroom up to a task-dependent quality ceiling; beyond it, only real bandwidth preserves throughput and answer quality simultaneously. Our measurement plan therefore tests utilisation and answer quality together (M5).

\subsection{Recompute, and the economics}
\begin{figure}[H]
\centering
\includegraphics[width=0.85\textwidth]{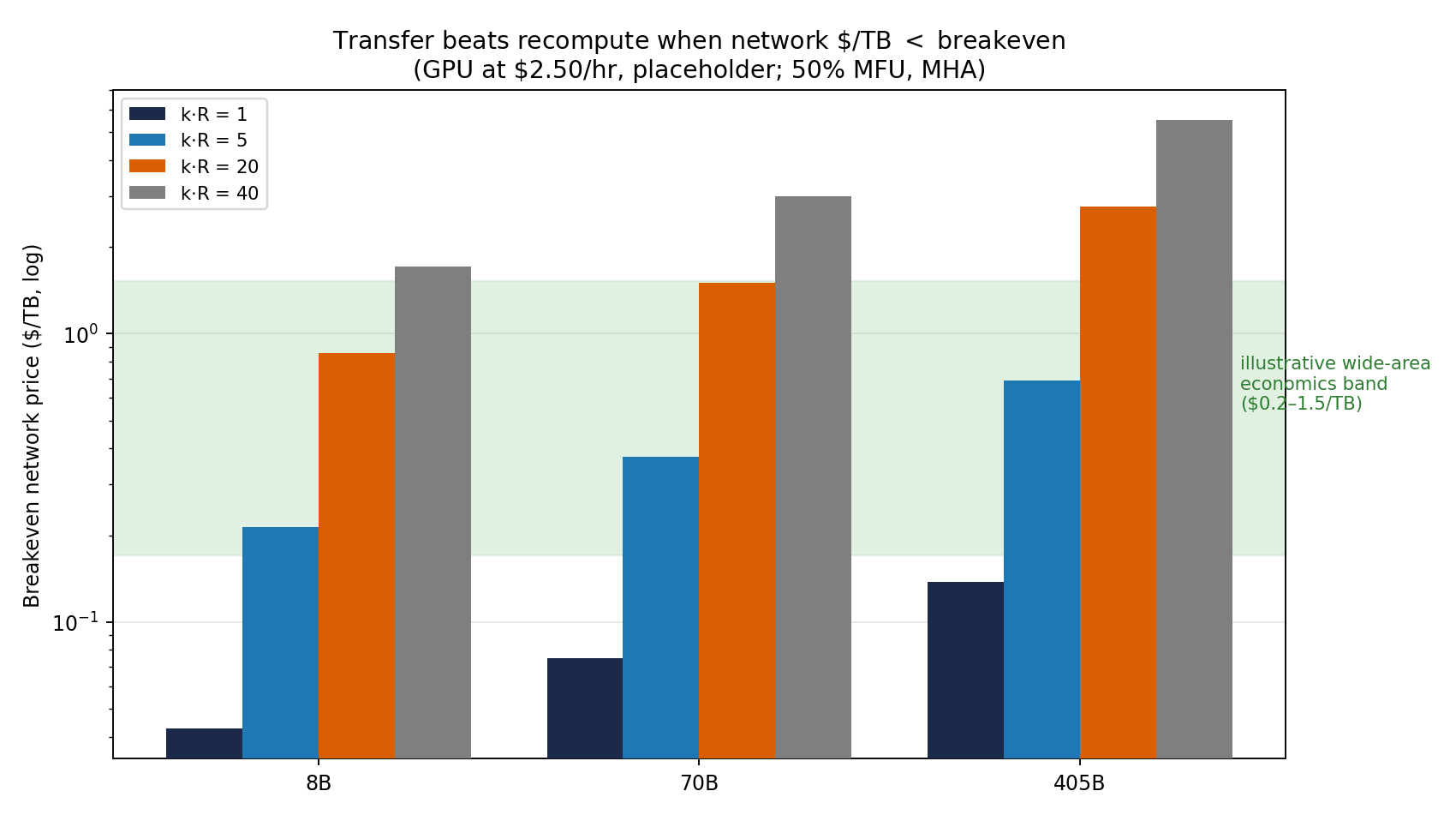}
\caption{Cost breakevens by model and multiplier $k \cdot R$ against an illustrative wide-area economics band (\$0.2--1.5/TB); at $k \cdot R = 1$ (MHA) recomputation wins, and modern attention or modest multipliers reverse it.}
\label{fig:cost}
\end{figure}

Cost parity between transfer and recompute is context-independent: at a GPU price of \$2.50/hr (a placeholder our measurement plan replaces), the breakeven network price is \$0.04--0.14 per TB moved (8B--405B, MHA). Effective wide-area economics run roughly \$0.2--1.5 per TB at realistic fill, so at face value recomputation wins and an honest cost model says so. Transfer wins through three multipliers: $k$---GPU scarcity (in a hot fleet a prefill GPU-second displaces revenue-bearing decode tokens, so effective GPU cost is marginal token revenue, not list price); $R$---KV reuse (one transfer substitutes $R$ recomputations across agentic steps and shared prefixes); and attention architecture---GQA moves 8$\times$ fewer bytes for the same avoided GPU-second, lifting the breakeven to $\sim$\$0.60/TB, which well-filled wide-area capacity already beats. The economic case for provisioned bandwidth is therefore regime-bounded: hot fleets, high reuse, modern attention---precisely GPU-scarce markets and agentic workloads. Figure~\ref{fig:cost} plots the breakevens against an illustrative wide-area economics band.

\subsection{Packet agility vs optical fungibility---allocation vs supply}
A fair comparison must include the network-side alternative to elastic optics: an agile packet fabric. Modern inference stacks are layer-3-and-above constructions---AI gateways, semantic classifiers, GPU-aware routers---and a well-run IP/SRv6 backbone reallocates traffic in milliseconds. The correct analysis is that packet and optical agility operate on different variables of the same supply chain. Statistical multiplexing, QoS, and traffic steering allocate capacity that is already lit; they cannot create it. When aggregate site-pair demand approaches lit capacity, queueing takes over regardless of routing intelligence: p99 queue wait grows from $\sim$5$\times$ the transfer time at 60\% utilisation to $>$20$\times$ at 90\% (\S3.4). At that point there are exactly two remedies: permanent headroom (overprovision) or a change in lit capacity (optical elasticity).

This reframes the thesis from ``optical is necessary'' to a falsifiable economic claim: optical fungibility substitutes for packet overprovisioning, not for packet agility. A hyperscaler overprovisioning its backbone 3--5$\times$ obtains every property described here with no visible optical layer---at hyperscaler capital intensity. For everyone else the comparison is the cost of static headroom (peak-to-mean $\times$ always-on \$/G) against on-demand capacity (duty-cycle $\times$ elastic \$/G, plus the control-loop latency to invoke it). A second, physical distinction: elephant-flow determinism. KV transfers are gigabyte-scale bursts occupying line rate for 0.1--1\,s; priority queueing protects small flows from large ones, not competing elephants from each other on a shared trunk. Sub-millisecond-variance delivery at high utilisation exists at layer 1---dedicated wavelengths or OTN channels---in a way no queueing discipline provides at layer 3 under load. The system consequence (\S4): the service interface is and should be packet---policy-classed L3---while elasticity and the determinism floor are supplied optically underneath.

\subsection{Concurrency and tails---the queueing knee}
\begin{figure}[H]
\centering
\includegraphics[width=0.85\textwidth]{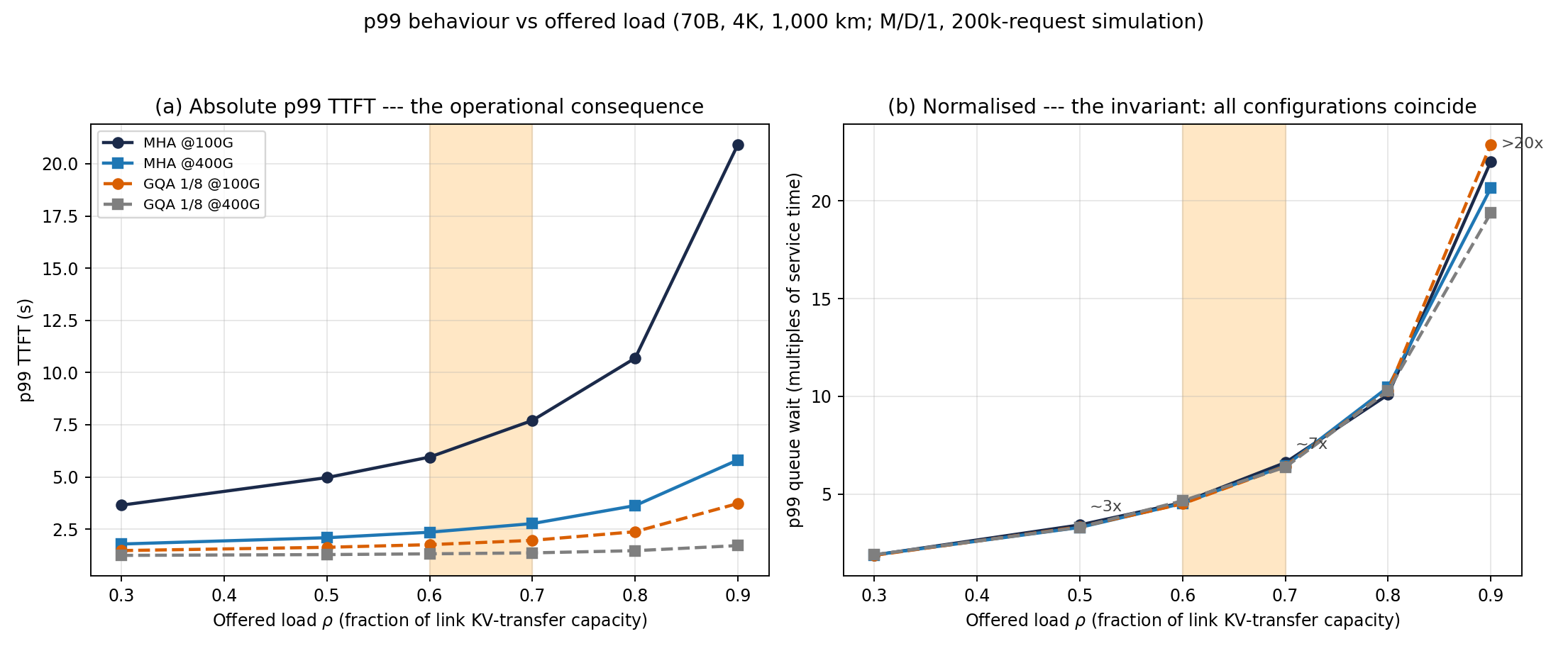}
\caption{(a) Absolute p99 TTFT: the operational consequence, which differs by an order of magnitude across configurations. (b) The same runs with queue wait normalised to service time: all four configurations coincide, because the p99 multiplier is workload-independent. The knee sits at $\rho \approx 0.7$ in both views.}
\label{fig:queue}
\end{figure}

Treating a site-pair link as an M/D/1 queue at the KV-transfer stage (Poisson arrivals, deterministic service): a 100G wave saturates at $\sim$1.16 70B-MHA requests/s ($\sim$0.8 req/s at the 70\% operating ceiling recommended below); 400G with GQA saturates at $\sim$37 req/s. The p99 queue wait in service-time units is workload-independent: $\sim$3$\times$ at 50\% utilisation, $\sim$7$\times$ at 70\%, $>$20$\times$ at 90\%---roughly doubling per +10 points past 70\%. Consequences: capacity planning for inference interconnect should treat $\sim$60--70\% utilisation as the operating ceiling, and any dynamic-capacity loop must trigger before the knee with headroom no smaller than its own provisioning latency (Figure~\ref{fig:queue}).

\subsection{Loss and jitter convert nominal bandwidth into much less effective bandwidth---for RDMA more than for TCP}
TCP first. Per-flow TCP-family throughput is bounded by $C \cdot \mathrm{MSS}/(\mathrm{RTT}\sqrt{p})$~\cite{mathis}; we take $C = 1.22 = \sqrt{3/2}$, the value for periodic loss with every packet acknowledged (delayed ACKs give 0.87, random loss 1.31, so the constants below shift by well under a factor of two under other assumptions). On measured production routes, with jumbo frames throughout: one flow at 56\,ms RTT and $10^{-5}$ loss caps at $\sim$0.5\,Gbps, and filling a 400G path needs $\sim$256 parallel flows even at $10^{-6}$ loss. Collectives amplify tails: a transfer scattered across $N$ flows completes at the maximum, so a per-flow p99 event lands in 47\% of 64-flow collectives ($1-0.99^{64}$), and $E[\max] \approx \sigma\sqrt{2\ln N} \approx 2.9\sigma$ at $N{=}64$. Parallelism thus fixes throughput and worsens tails simultaneously.

RDMA is the sharper case---and AI infrastructure prefers RDMA. Classic RoCEv2 was designed for lossless fabrics: its congestion control (DCQCN) reacts to ECN marks at microsecond RTTs, not loss, and its loss recovery is go-back-N---one lost packet forces re-transmission of roughly the entire in-flight window, and its NICs treat out-of-order arrival as loss~\cite{irn}. Neither assumption survives the WAN: PFC pause feedback at 56\,ms arrives $\sim$2.8\,GB of in-flight data too late, and the in-flight window at 400G$\times$56\,ms is $\sim$311{,}000 packets. A first-cut efficiency model (goodput $\approx 1/(1 + p \cdot W)$, $W$ = in-flight packets; deliberately generous---it ignores timeouts and concurrent congestion response) gives Table~\ref{tab:transport}:

\begin{table}[H]
\centering\small
\caption{First-cut go-back-N goodput model for RoCEv2 at WAN distances, 9\,kB frames. Deliberately generous: it ignores timeouts and concurrent congestion response. Offered as observations for industry discussion, not as measurements.}
\label{tab:transport}
\begin{tabular}{llllll}
\toprule
link, RTT & $W$ (pkts) & goodput @$10^{-6}$ & @$10^{-5}$ & loss for $\geq$90\% goodput & reordering tolerance\\
\midrule
100G, 32 ms & 44{,}444 & 96\% & 69\% & $p \leq 2.5\times10^{-6}$ & $\approx$ 0 (OOO = loss)\\
100G, 56 ms & 77{,}778 & 93\% & 56\% & $p \leq 1.4\times10^{-6}$ & $\approx$ 0 (OOO = loss)\\
400G, 56 ms & 311{,}111 & 76\% & 24\% & $p \leq 3.6\times10^{-7}$ & $\approx$ 0 (OOO = loss)\\
400G, 114 ms & 633{,}333 & 61\% & 14\% & $p \leq 1.8\times10^{-7}$ & $\approx$ 0 (OOO = loss)\\
\bottomrule
\end{tabular}
\end{table}

\begin{figure}[H]
\centering
\includegraphics[width=0.85\textwidth]{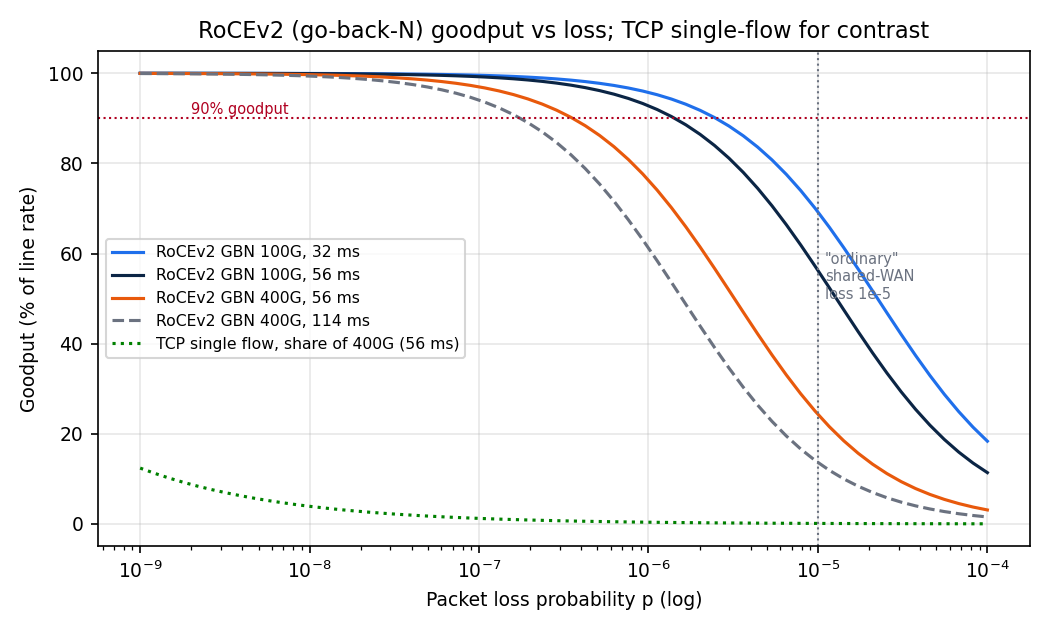}
\caption{RoCEv2 (go-back-N) goodput vs loss at WAN distances; single-flow TCP shown for contrast.}
\label{fig:transport}
\end{figure}

Readings. At ``ordinary'' shared-WAN loss ($10^{-5}$), RoCEv2 at 400G/56\,ms collapses to $\sim$24\% goodput---the pipe stays busy, mostly re-sending. Holding $\geq$90\% goodput requires paths roughly one to two orders of magnitude cleaner than a shared WAN, and the reordering tolerance is effectively zero: delay variation that reorders packets (per-packet spraying, unequal parallel paths) is loss-equivalent, so the requirement is a single deterministic path per connection with delay variation well under the retransmit-timer floor. Selective-repeat NICs~\cite{irn} repair the waste but leave a congestion loop tuned for microsecond feedback, with no validated model at WAN RTT---which is precisely the gap that distance-aware congestion control efforts in the GPU-vendor ecosystem now target~\cite{xgs}, and independent confirmation that scale-across transport is an unsolved, actively-invested problem. The asymmetry is the finding (Figure~\ref{fig:transport}): TCP degrades gracefully on dirty paths; RDMA-class transports effectively require clean ones. A loss/jitter floor is therefore a bandwidth specification, not a quality garnish, and per-flow tail commitments must be engineered so that the per-flow exceedance probability is $\sim 1/N$ of the customer-visible target---the quantitative case for deterministic layer-1 delivery classes, made strongest by the transport the AI stack actually uses.

\subsection{A third option the ecosystem should quantify: storage-tier reload}
Between transfer and recompute sits a third option this paper has not modelled: checkpoint the KV to storage (local NVMe or object store) and reload on demand. It is cheap per byte and avoids both the WAN and the GPU---at the price of cold-start latency, capacity management, and staleness for growing agentic contexts. A complete design-space treatment should bound it; we flag it as open work and an invitation---this is exactly the kind of contribution an ecosystem effort should absorb rather than one operator model in isolation.

\section{System design: a two-timescale control plane}
The architecture follows from \S3.3. Per-request decisions (which site, which policy class) execute in milliseconds at layers 3--7: gateway classification, GPU- and KV-aware inference routing, policy-classed L3 VCs as the customer-visible service unit. Capacity decisions (how much lit bandwidth a site-pair has) execute in seconds-to-minutes in the optical NaaS layer: client-port capacity steps on pre-lit wavelengths, new-wavelength turn-up, protection. Whatever switching technology supplies the second loop, it is never in the per-request path: capacity is changed on the seconds-to-minutes timescale of provisioning, protection, and topology, not on the timescale of a request. The loop closes on an explicit trigger: the inference layer requests capacity when site-pair utilisation crosses $\sim$0.6--0.7 (the \S3.4 knee), with the threshold set below the knee by at least the measured allocation-loop latency. Aggregate demand---never per-request events---drives capacity: a single prompt moves kilobytes; wavelengths scale on site-pair volume crossing thresholds.

\section{Methodology: an open measurement plan for the ecosystem}
The analytical model makes falsifiable predictions. Measuring them end to end requires a testbed no single company can---or should---assemble alone: production wide-area fibre with an elastic optical layer, a modern L3 fabric, and a real inference serving stack generating RDMA-class state transfer over actual wavelengths. We are assembling the network half on a new deep-buried production corridor in India: three sites, open optical line systems from multiple vendors under interoperability test, an SDN-driven optical control plane, disaggregated prefill/decode serving with synthetic transfer-path load generators standing in for GPU workers and a scheduler simulator for compute-side sweeps, hardware load generation, and TWAMP delay/jitter measurement.

The measurement plan is deliberately framed as an open ecosystem exercise, with the division of labour following the stack. The operator measures what only an operator can: provisioning and capacity-step latency, delay-variation and protection percentiles, allocation-loop timing, multi-tenant isolation, and unit cost. Optical and packet vendors bring line systems and fabrics into interoperability test. The inference ecosystem---GPU vendors, serving-stack and inference-fabric providers---brings the workload truth: measured TTFT and utilisation, the real transfer-vs-recompute flip point, compression-quality curves, locality behaviour, and above all an RDMA-class transport running over real wavelengths, without which every number in \S3.5 remains analytic. Ten metrics, each tied to a prediction (Table~\ref{tab:metrics}):

\begin{table}[H]
\centering\small
\caption{The ten-metric measurement plan, with the owner class and the specific model prediction each metric tests.}
\label{tab:metrics}
\begin{tabular}{lp{7.2cm}lp{4.2cm}}
\toprule
\# & Metric & Owner class & Tests\\
\midrule
M1 & Wave provisioning + capacity-step time & operator & control-loop supply latency\\
M2 & p50/p95/p99 PDV + protection-switch behaviour & operator & \S3.5 tail assumptions\\
M3 & Measured TTFT \& GPU utilisation, 100G vs 400G & inference ecosystem & predicts 57\%/84\% util (70B-MHA, 4K)\\
M4 & Transfer-vs-recompute flip point at swept link rates & inference ecosystem & predicts 74--111 Gbps MHA; 9--14 Gbps GQA-1/8\\
M5 & Compression: utilisation AND answer quality at 1$\times$/4$\times$/8$\times$ & inference ecosystem & predicts 12$\to$35$\to$52\% util; quality ceiling~\cite{memento}\\
M6 & Cross-site demand vs prefix-cache hit rate & inference ecosystem & predicts demand $\propto$ (1 $-$ hit rate)\\
M7 & Dynamic-allocation loop latency end-to-end & operator + partner & \S4 trigger headroom\\
M8 & Multi-tenant isolation under adversarial burst & operator & \S3.3 elephant-flow claim\\
M9 & TTFT p50/p99 under concurrency sweep & operator + partner & \S3.4 M/D/1 curves\\
M10 & Measured \$/G and effective \$/TB at pilot scale & operator & replaces \S3.2 placeholders\\
\bottomrule
\end{tabular}
\end{table}

The pilot window targets late 2026 and is open to industry collaborators; partner-owned metrics are reported only if measured by the owning partner, and are otherwise marked future work. The plan includes an RDMA-over-wavelength trial where NIC support permits.

\section{Analytical evaluation (pre-measurement)}
Baseline: the model reproduces our published India simulation~\cite{lightstorm} within rounding (9.75\,s TTFT / 12\% GPU utilisation at shared 10G; 1.38\,s / 84\% at 400G; 70B, 4K, 1{,}000\,km). Context: KV is linear in tokens, so a 128K context needs $\sim$44\,s to first token even at 400G ($\sim$37\,s prefill plus $\sim$7\,s transfer)---interactive long-context disaggregation is out of scope for the WAN regardless of supply, and compute rather than transport is the binding term. Model size: larger models tolerate the WAN better (prefill compute grows faster than KV): 8B/70B/405B at 400G give 73/84/91\% utilisation. Distance: for relaxed TTFT targets propagation is $<$2\% of TTFT even at 4{,}000\,km---bandwidth-bound at all distances; for a 40\,ms interactive target, physics closes the door beyond metro radius and no bandwidth reopens it. What weakens our own case, stated plainly: GQA/MLA short-context serving with high locality needs neither large pipes nor optical elasticity; and at list GPU prices with no reuse, recomputation beats transfer everywhere. The thesis survives where fleets are hot, KV is reused, contexts are long, sessions are agentic, and tails are contractual---and we claim it only there.

\section{Discussion}
Reliability and isolation are measured (M2, M8), not asserted. The SLA structure that follows from \S3.5 is two-regime: a committed steady-state p99 delay-variation floor per delivery class, plus separately disclosed protection-event behaviour---steady-state variance determines effective bandwidth; events determine availability of the deterministic class. Limits of the analysis: Poisson/FIFO arrivals (real gateways batch and preempt); placeholder dollar inputs until M10; the locality fraction is free until M6 binds it; the loss model is Reno-shaped and modern congestion control shifts constants; storage-tier reload (\S3.6) is unmodelled. Each limitation maps to a named metric---that mapping is the point.

\section{Conclusion}
The CIR gap is structural, and its shape is the argument. On a consistent dense basis the B200 generation holds on-package CIR roughly flat (HBM $\sim$3.56\,mB/F at 2{,}250 TFLOPS dense BF16, against H100's 3.39 at 989) and lets scale-up slip only modestly (NVLink $\sim$0.80\,mB/F against 0.910): the tiers the industry has funded have broadly tracked compute. The wide-area tier has not, and remains the one segment of the GPU-to-GPU continuum without an investment thesis attached. The correct engineering response is not ``add optical bandwidth everywhere.'' It is a co-design with an explicit decision boundary: recompute below the crossover; exploit locality and compression up to the quality ceiling; serve short-context GQA traffic on ordinary packet capacity; and supply hot-fleet, high-reuse, long-context, and agentic regimes with elastic provisioned bandwidth---packet-fronted at the service layer, optically elastic underneath, triggered at the queueing knee, with tails engineered at layer 1. Every element of that sentence is measurable, and \S5 is the open plan to measure it. An industry effort built on this evidence should say to the market what Metro Ethernet's requirements said two decades ago: here is the missing tier, here is how to specify it, and here is the evidence.

\section*{Acknowledgements}
The author used AI-assisted tools (Anthropic's Claude) in the preparation of this work: implementation of the reproducible sensitivity model and discrete-event simulations, drafting and editorial support, and literature retrieval. All analytical framings, design decisions, and conclusions are the author's. Every quantitative claim derives from the accompanying model and was reviewed by the author, and all references were verified by the author against their original sources.

\end{document}